\documentclass[12pt,cite,a4paper]{article}

\usepackage{amsmath}
\usepackage{amssymb}
\usepackage{a4wide}
\usepackage{amssymb}
\usepackage{graphicx}
\usepackage{mathrsfs}

\newcommand{\cR}{\mathcal{R}}
\newcommand{\cT}{\mathcal{T}}
\newcommand{\be}{\begin {equation}}
\newcommand{\ee}{\end {equation}}
\newcommand{\ba}{\begin{eqnarray}}
\newcommand{\ea}{\end{eqnarray}}

\def\Im{\mathop{\rm Im}}
\def\Re{\mathop{\rm Re}}
\def\bndy#1{\mbox{\scriptsize ${\star \atop\star}$} #1 \mbox{\scriptsize${\star\atop \star}$}}
\numberwithin{equation}{section}

\begin{document}
%\rightline{THU-IAS-09-XX}
 \vspace{2truecm}

%%%%%%%%%%%%%%%%%
\centerline{\Large \bf Note on Light-like Tachyon Condensation}

\vspace{2.3truecm}

\centerline{
    {Yushu Song }%\footnote{Electric address: yssong@mail.tsinghua.edu.cn}
    }
\vspace{.8cm} \centerline{{\it Institute for Advanced Study, Tsinghua University, Beijing, China, 100084}}
\vspace*{2.0ex} \centerline{E-mail: {\tt yssong@mail.tsinghua.edu.cn}}

\vspace{2.5truecm}

%%%%%%%%%%%%%%%%%
\centerline{\bf ABSTRACT}
\vspace{.5truecm}

In this paper closed string emission and open string pair production from the light-like rolling tachyon solution are calculated in subcritical string theory in the background of a linear dilaton. The rolling light-like tachyon represents the inhomogeneous decay of unstable D-brane. The decay rate is given by the imaginary part of annulus diagram which can be calculated using the boundary state/$\sigma$-model method. It is found that the decay rate is finite in the open string ultraviolet region and depends on tachyon profile in the open string infrared region.

 \noindent

%%%%%%%%%%%%%%%%%%%%%%%%%%%%%%%%%%%%%%%%%%%%%%%%%%%%%%%%%
\newpage

\section{Introduction}

It is of great interest to understand the dynamical process of string theory. One of the most efficient ways to generate time dependent configurations in string theory is to consider the decay of unstable D-branes. In Bosonic string theory D$_p$-branes  are unstable due to the existence of the tachyon in the open string spectrum. Tachyon condensation is pictured as a tachyon field rolling down a potential towards a stable minimum which is the closed string vacuum \cite{Sen:1999mg}. The open string tachyon condensation, initiated by Sen \cite{Sen:2002nu}-\cite{Sen:2002an}, has been intensively studied in the past decade. For a comprehensive review of open string tachyon dynamics, see \cite{Sen:2004nf}.\\

Until now the process of time-like tachyon condensation is quite well-studied with the help of boundary state/$\sigma$-model method. This method is based on the well-known correspondence between classical solutions of equations of motion for string theory and two dimensional conformal field theories. The classical picture of D-brane decay in which the string coupling constant is strictly vanishing, shows that the late time evolution of the tachyon leads to a pressureless fluid which is called ``tachyon matter''. This seems strange because if the tachyon fully condenses, the D-brane should disappear and there should be no open string excitations.  All excitations should be described in terms of closed string. In order to answer the question whether the tachyon matter is closed string in disguise or a new degree of freedom, it is natural to consider the D-brane decay with the nonvanishing string coupling constant \cite{Chen:2002fp} \cite{Lambert:2003zr}. The D-brane with a rolling tachyon acts as a linear source, a tadpole, for each closed string mode. The radiation can be captured by one point function of the closed string vertex operator on the disc. Detailed calculations demonstrate that the energy of emitted radiation from unstable D$_p$-brane diverges for $p=0,1,2$, and is formally finite for $p>2$. When considering the closed string emission rate in Bosonic string theory, authors of \cite{Lambert:2003zr} found a Hagedorn-like divergence which was argued to lead to the breakdown of perturbation theory. Thus we avoid the direct confrontation with tachyon matter. It was further suggested \cite{Lambert:2003zr} \cite{Sen:2003bc} that the tachyon matter might be composed of highly massive closed strings. Subsequently when the above analysis was generalized to the subcritical string theory in a space-like linear dilaton background \cite{Karczmarek:2003xm}, some interesting results were obtained. The expression for closed string radiation is finite. In this setting perturbation theory does not seem to break down, and the tachyon matter drifts into the weak coupling region. Thus the tachyon matter puzzle seems to reappear. More general background was discussed in \cite{He:2006bm}.\\

Most of the developments described above are based on time-like tachyon profile, which only depends on the time direction\footnote{For inhomogenous tachyon condensation, see \cite{Sen:2002vv} \cite{Larsen:2002wc}.}. On the other hand, light-like tachyon solution was constructed in \cite{Hellerman:2006ff} \cite{Hellerman:2006hf} in the context of closed string tachyon condensation. Some aspects of light-like tachyon condensation were analyzed in the framework of open string field theory \cite{Hellerman:2008wp}. Compared with time-like tachyon profile, light-like tachyon is more physically relevant since for the tachyon profile with spatial momentum $\vec{k}$ and effective mass $m^2=-1+\vec{k}^2$, all wave lengths play some role in the brane decay process. Light-like tachyon condensation describes inhomogeneous decay of unstable D-brane which is a more general process than homogenous decay of unstable D-brane described by time-like tachyon condensation. In this paper, the light-like tachyon condensation is analyzed. The worldsheet partition function for unstable D-brane decay is calculated. As a result of optical theorem, the imaginary part of the partition function is interpreted as the closed string emission which is consistent with open string pair production. This paper is organized as follows. In section 2, we review some facts about string theory in the linear dilaton background. In section 3, we compute the annulus diagram using boundary state/$\sigma$-model method, and also analyze the asymptotic behavior of the imaginary part of the annulus diagram in some region of the parameter space. In section 4, we calculate the open string pair production and analyze its asymptotic behavior. In appendix A, we perform the detailed calculation of exact bulk one-point function.

%\newpage
%%%%%%%%%%%%%%%%%%%%%%%%%%%%%%%%%%%%%%%%%%%%%%%%%%%%%%%%%%%%%%%
%%%%%%%%%%%%%%%%%%%%%%%%%%%%%%%%%%%%%%%%%%%%%%%%%%%%%%%%%%%%%%%
\section{Strings in linear dilaton background}
String theory in flat space with a linear dilaton  $\Phi=V_\mu X^\mu$ is based on the action \cite{Polchinski:1998rq}
\be \label{ld}
S_{\rm LD}=\frac{1}{4\pi\alpha'}\int_\Sigma d^2\sigma\sqrt{g}(g^{\alpha\beta}\partial_\alpha X^\mu\partial_\beta X_\mu+\alpha'\cR V_\mu X^\mu)+\frac{1}{2\pi}\int_{\partial\Sigma}ds KV_\mu X^\mu
\ee
where $g_{\alpha\beta}$ is the intrinsic worldsheet metric in Euclidean signature, $\cR$ is the Ricci scalar of worldsheet $\Sigma$, and $K$ is the extrinsic curvature of boundary $\partial{\Sigma}$. With the diffeomophism and Weyl invariance fixed (assuming the central charge to be zero with the ghost sector included), the metric can be set to the canonical form $ds^2=dzd\bar{z}$.
The energy momentum tensor from action (\ref{ld}) is given as follows\footnote{From here on, we take the convention $\alpha'=1$.}
\ba
T(z)=-:\partial X^\mu\partial X_\mu:+V_\mu\partial^2 X^\mu \\
T(\bar{z})=-:\bar{\partial} X^\mu\bar{\partial} X_\mu:+V_\mu\bar{\partial}^2 X^\mu
\ea
The corresponding Virasoro generators are modified to be
\be
L_m^{\rm matter}=\frac{1}{2}\sum^\infty_{n=-\infty}: \alpha_{m-n}^\mu \alpha_{\mu n}:
+  \frac{i}{\sqrt{2}} (m+1) V^\mu \alpha_{\mu m}
\ee
The central charge is modified to be $c=\tilde{c}=D+6V^2=26$, where $D$ is the number of spacetime dimensions and $V^2=V_\mu V^\mu$. From the operator product expansion with energy momentum tensor, conformal weight of all primary operators can be determined. For example, the weight of bulk operator $:e^{i k X(z, \bar z)}:$ is
\be
h=\frac{k^2}{4} + i \frac { V^\mu k_\mu}{2}
\ee
and the weight of boundary operator $\bndy{e^{i k
X(y)}} $ is
\be
h=k^2+ i k_\mu V^\mu
\ee
The on-shell condition for closed strings is
\be
L^{\rm closed}_0 \mid {\rm phys}> =(\frac{P^2}{4}+N+\frac{i}{2}V_\mu P^\mu -1) \mid {\rm phys} >=0
\ee
which gives the physical state condition for bulk operators.
The on-shell condition for open strings is
\be
L^{\rm open}_0 \mid {\rm phys}> =(P^2+N+iV_\mu P^\mu -1) \mid {\rm phys} >=0
\ee
which gives the physical state condition for boundary operators.
Then for the open string light-like tachyon profile $\cT=\exp bX^+$, the corresponding vertex operator introduces the worldsheet boundary interaction
\be
S_{\rm int}=\frac{\lambda}{2\pi}\int_{\partial \Sigma}ds e^{b X^+}
\ee
which describes the unstable brane at early $X^+$ and its decays.
The physical condition for open string tachyon state results in $bV^+=1$ which is just the marginal condition for the boundary interaction induced by the tachyon vertex operator. In order to avoid the force on the brane, dilaton gradient is required to point along the unstable brane.  We will work in the light-cone coordinates
\ba
X^{\pm}&=&\frac{1}{\sqrt{2}}(X^0 \pm X^1) \\
ds^2&=&-2dX^+dX^-+dX^2_2+...+dX^2_{D-1}
\ea
About the path integral formulism in linear dilaton background, there is a systematic review in the appendix of \cite{Ho:2007ar}. For later convenience, we will review some known results of the path integral formulism in the linear dilaton background. The linear dilaton background modifies the correlation functions in two ways. First, it changes the worldsheet boundary condition for the open string. When we derive the equations of motion from action (\ref{ld}), the surface term from variation is
\ba
\left(\partial_n X^\mu+K V^\mu \right)\delta X_\mu=0~~~~~\rm{on}~~~ \partial\Sigma
\ea
where $n$ is the normal vector of the boundary $\partial\Sigma$. Second, it modifies the momentum conservation law. Consider the simplest boundary correlators
\ba
\langle e^{ik_1X(y_1)}e^{ik_2X(y_2)}...e^{ik_nX(y_n)} \rangle_{\Sigma}=iC_{\Sigma}^X(2\pi)^D\delta^D(\Sigma k^{\mu}_i+i\chi_{\Sigma}V^{\mu})\prod_{1\leq i< j\leq n}|y_i-y_j|^{2k_ik_j}
\ea
where $C_{\Sigma}^X$ is some constant
\ba
C_{\Sigma}^X=\left(\int_{\Sigma}d^2\sigma g^{1/2}\right)^{D/2}\left({\det}^{\prime}\frac{-\nabla^2}{4\pi^2} \right)^{-D/2}
\ea
and $\chi_{\Sigma}$ is the Euler character of the worldsheet $\Sigma$
\ba
\chi_{\Sigma}=\frac{1}{2\pi}\int_{\partial\Sigma}d s K +\frac{1}{4\pi}\int_{\Sigma}d^2\sigma g^{1/2}\cR ~~~.
\ea

\section{Closed string emission}
In this section, we discuss the D-brane decay process in a linear dilaton background. We will compute the partition function in the closed string channel, extract its imaginary part and discuss the various divergences in some asymptotic limits.

\subsection{Partition function}
The open string partition function in the closed string channel can be written as
\be \label{tpf}
Z(\tilde{q})=\langle B |\tilde{q}^{L_0-\frac{D-2}{24}}|B\rangle
\ee
where  $\langle B|$, the boundary state, is a closed string state of ghost number 3, and has the following properties. With a closed string state $ | V\rangle $ and the associated vertex operator $V$ in hand, the BPZ inner product $\langle B|V\rangle $ is given by one-point function of $V$ inserted at the center of a unit disc $D$. The boundary condition on $\partial D$ is the one associated with the particular boundary CFT under consideration:
\ba
\langle B|V\rangle  \varpropto  \langle V(0) \rangle_D
\ea
The expression (\ref{tpf}) can be factorized into light-cone directions part and space-like directions part. We will first study the light-cone directions part in detail. Following the procedure developed in \cite{Teschner:2000md}, we can write down the light-cone directions partition function
\ba
&Z^\pm(\tilde{q})=\langle B_\pm |\tilde{q}^{L_0-\frac{D-2}{24}}|B_\pm \rangle & \label{ccz}
\ea
where $\langle B_\pm |$ is the boundary state of $X^\pm$ directions. According to the definition of the boundary state specified above  $\langle B_\pm |$ can be written as
\ba
&\langle B_\pm |=\int_0^\infty \frac{d\omega}{2\pi}\frac{d\omega'}{2\pi}~U(\omega,\omega')\langle \omega |\otimes\langle \omega' | & \label{bs}
\ea
where $\langle \omega |$ and $\langle \omega' |$ are the so-called Ishibashi states \cite{Ishibashi:1988kg}. $U(\omega,\omega')$ is the one point function of  $e^{\alpha_+X^++\alpha_-X^-}$ on the disc. The operator $e^{\alpha_+X^++\alpha_-X^-}$ should be dressed, so $\alpha_+=V_+-i\omega$ and $\alpha_-=V_--i\omega'$. Combining (\ref{ccz}) and (\ref{bs}), we have
\be
Z^\pm(\tilde{q})=\int\frac{d\omega}{2\pi}\frac{d\omega'}{2\pi}|U(\alpha_+,\alpha_-)|^2\chi_{\alpha_+}(\tilde{q}) \chi_{\alpha_-}(\tilde{q})
\ee
where $\chi_{\alpha_\mu}$ is the character
\be
\chi_{\alpha_\mu}(q)=\eta^{-1}(q)~q^{\frac{(\alpha_\mu-V_\mu)^2}{4}}
\ee
The closed string channel parameter $\tilde{q}$ and open string channel parameter $q$ are defined as
\ba
&\tilde{q}=\exp(-2\pi i/\tau)=e^{-4t}& \\
&q=\exp(2\pi i\tau)=e^{-\frac{\pi^2}{t}}&
\ea
As for the calculation of one point function $U(\alpha_+,\alpha_-)$, we will use the technique developed in \cite{Hellerman:2008wp} to work out the integral. In order to deal with the boundary term, Hellerman and Schnabl fix the value of zero mode $x^\mu$ of the embedding coordinate $X^\mu$ as a degree of freedom in the boundary state wave-function. From the path integral point of view, this amounts to fixing
\be
x^\mu\equiv x^\mu_{\rm boundary}\equiv\frac{1}{2\pi}\int_{\partial D}dsX^\mu
\ee
In our case, we will take $V_+=0$ for convenience. Thus we do not need to fix the zero mode of $X^+$.
About directions whose zero modes are fixed, the propagator is modified as follows
\be
\left\langle X^\mu(z,\bar{z}) X^\nu(z',\bar{z}') \right\rangle=x^\mu x^\nu+\eta^{\mu\nu}P(z,\bar{z};z',\bar{z}')
\ee
where
\be \label{propagator}
P(z,\bar{z};z',\bar{z}')=-\frac{1}{2}\left[\ln|z-z'|^2 + \ln|1-z\bar{z}'|^2 \right]
\ee
on the unit disc. Before continuing our discussion, we should make a clarification on two kinds of normal-orderings. The bulk normal-ordering, denoted here by $:~~:$, subtracts only the first logarithm in (\ref{propagator}) which is
\ba
&:X^\mu(z,\bar{z}) X^\nu(z,\bar{z}'):\equiv X^\mu(z,\bar{z}) X^\nu(z',\bar{z}')+\frac{1}{2}\eta^{\mu\nu}\ln|z-z'|^2&
\ea
The boundary normal-ordering, denoted here by $\mbox{\scriptsize ${\star \atop\star}$}~~\mbox{\scriptsize ${\star \atop\star}$}$, subtracts the full propagator (\ref{propagator}) which is
\ba
&\bndy{ X^\mu(z,\bar{z}) X^\nu(z,\bar{z}')}\equiv X^\mu(z,\bar{z}) X^\nu(z,\bar{z}')+\frac{1}{2}\eta^{\mu\nu}\left[\ln|z-z'|^2 + \ln|1-z\bar{z}'|^2 \right] &
\ea
In the following subsection, we will calculate the one-point function of the closed string vertex operator $V_{\alpha_+,\alpha_-}=e^{\alpha_+X^+ +\alpha_-X^-}$ on a unit disc $D$. Since zero mode of $X^+$ direction is not fixed, the one-point function can be written as
\ba \label{one-point}
\langle V_{\alpha_+,\alpha_-}\rangle=\int dx^+dx^-D\hat{X}^+D\hat{X}^-\delta\left(x^--\frac{1}{2\pi}\int_{\partial D}dsX^-\right)e^{-S^\pm}e^{\alpha_+X^+ +\alpha_-X^-}
\ea
where
\be
S^\pm=-\frac{1}{2\pi}\int_D d^2\sigma\sqrt{g}g^{\alpha\beta}\partial_\alpha X^+\partial_\beta X^-+\frac{1}{2\pi}\int_{\partial D}ds(V_- X^-+ \lambda e^{b X^+})
\ee
Performing the perturbative expansion of the one-point function (\ref{one-point}), we have
\ba \label{one-point2}
\langle V_{\alpha_+,\alpha_-}\rangle=\int dx^+\sum_{n=0}^\infty \frac{(-\lambda)^n}{n!}\left(\prod_{i=1}^n\int_0^{2\pi}\frac{dt_i}{2\pi} \right)\langle :e^{\alpha_+X^+ +\alpha_-X^-}(z,\bar{z}): \prod_{i=1}^n \bndy{e^{b X^+[\omega_i(t_i)]}}\rangle_{\rm LD}\nonumber\\
\ea
where the brackets on the right hand side represent the expectation value calculated in the linear dilaton background. With the vertex operator placed at the origin, the one-point function can be carried out from (\ref{one-point2})
\ba
U(\alpha_+,\alpha_-)&=&\sum_{n=0}^\infty \frac{(-\lambda)^n}{n!} \int^{+\infty}_{-\infty}  dx^+ \exp{\left[(\alpha_+-V_+)x^++(\alpha_--V_-)x^- +nbx^+\right]}\nonumber \\
&=&\frac{\lambda^{\frac{i\omega}{b}}}{b}e^{-i\omega'x^-}\Gamma(-\frac{i\omega}{b})
\ea
and we obtain
\be \label{usquare}
|U(\alpha_+,\alpha_-)|^2=\frac{\pi}{b\omega\sinh\frac{\pi\omega}{b}}
\ee
From formula (\ref{usquare}), we find that $|U(\alpha_+,\alpha_-)|^2$ only contains information from $X^+$ direction. $X^-$ which decouples from $X^+$ direction, has the same form $|U|^2$ as other directions, whose partition function can be calculated using the standard method. Now we only need to calculate the partition function of $X^+$ direction, which can be written as
\be \label{z+}
Z^+(\tilde{q})=\int\frac{d\omega}{2\pi}|U(\alpha_+,\alpha_-)|^2\chi_{\alpha_+}(\tilde{q})
\ee
The oscillating closed string characters can be expressed as a modular transformation of non-oscillating open string characters \cite{Karczmarek:2003xm}\footnote{We would like to thank the authors of \cite{Karczmarek:2003xm} for explaining the details on this point.}
\be \label{chi}
\chi_{\alpha_+}(\tilde{q})=\sqrt{2}\int_{-\infty}^{+\infty}d\nu \cosh(2\pi\omega\nu)\chi_{\alpha'_+}(q)
\ee
where $\alpha'_+=2\nu+iV_+$. With the help of equation (\ref{chi}), $X^+$ direction partition function can be reduced to the following form
\be
Z^+(\tilde{q})=\sqrt{2}\int_{-\infty}^{+\infty}d\omega d\nu \frac{\cosh(2\pi\omega\nu)}{2\omega b\sinh\frac{\omega}{b}}\chi_{\alpha'_+}(q)
\ee
This expression has a double pole at $\omega=0$, which can be regulated by a subtraction following the procedure of \cite{Teschner:2000md} \cite{Karczmarek:2003xm}
\be \label{z+integral}
Z^+(\tilde{q})=\sqrt{2}\int_{-\infty}^{+\infty}d\omega d\nu \left(\frac{\cosh(2\pi\omega\nu)}{2\omega b\sinh\frac{\omega}{b}}-\frac{1}{2\omega^2}\right)\chi_{\alpha'_+}(q)
\ee
It is hard for us to work out this integral and we will analyze this integral in the limit $b\rightarrow 0$ and $b\omega\rightarrow 0$. The limit describes the tachyon profile that varies slowly along the $X^+$ compared with itself. With reference to the result of \cite{Karczmarek:2003xm}, the integral (\ref{z+integral}) can be expressed as
\be \label{mr}
Z^+(\tilde{q})=2\sqrt{2}\pi i\eta^{-1}(q)\sum_{n,m=0}^{\infty}e^{-[(n+\frac{1}{2})b+(m+\frac{1}{2})/b]^2\frac{\pi^2}{t}} \ee
Notice the formula (\ref{mr}) is valid only if the limits $b\rightarrow 0$ and $b\omega\rightarrow 0$ are taken on both sides. We will not emphasize this point in the following formulas and conclusions. The other part of partition function is
\ba
Z^{X^-+X^i+bc}&=&V_p\int\frac{d^{D-1-p}k_\perp}{(2\pi)^{D-1-p}}\tilde{q}^{\frac{k^2_\perp}{4}}\eta^{3-D}(\tilde{q})\nonumber \\
&=&\frac{\mathcal{N}^{-2}_pV_p}{2\sqrt{2}\pi t}\int\frac{d^pk_\parallel}{(2\pi)^p}q^{k^2_\parallel}\eta^{3-D}(q)
\ea
where $\mathcal{N}_p$ is the normalization of the boundary state
\be \label{nfn}
\mathcal{N}_p=\pi^{\frac{D-4}{4}}(2\pi)^{\frac{D-2}{4}-p}
\ee
Putting them together, we have
\ba \label{mr1}
Z(q)&&=\mathcal{N}^2_pZ^{X^-+X^i+bc}Z^+ \nonumber \\
&&=\frac{iV_p}{t}\sum_{n,m=0}^{\infty} \int\frac{d^pk_\parallel}{(2\pi)^p}q^{k^2_\parallel}\eta^{2-D}(q) e^{-[(n+\frac{1}{2})b+(m+\frac{1}{2})/b]^2\frac{\pi^2}{t}}
\ea
The imaginary part is
\be \label{imfp}
\Im Z(q)=\sum_{n,m=0}^{\infty}\frac{V_p}{t}\left(\frac{t}{4\pi^3} \right)^{\frac{p}{2}}\eta^{2-D}(q) e^{-[(n+\frac{1}{2})b+(m+\frac{1}{2})/b]^2\frac{\pi^2}{t}}
\ee
Though our main results (\ref{mr1}) and (\ref{imfp}) are only valid in the limit specified above, we can analyze its asymptotic behaviors and obtain some interesting results.

\subsection{Closed string radiation and its divergences}

Performing a modular transform on (\ref{imfp}) and integrating over $s=\pi^2/t$, we have
\ba
\Im\mathbf{Z}&=&\Im \int ds Z\left(\frac{is}{2\pi} \right) \nonumber \\
&=&V_p\sum_{n,m=0}^{\infty}\int_0^{\infty}\frac{ds}{s}\frac{\eta^{2-D}(\frac{is}{2\pi})}{(4\pi s)^{\frac{p}{2}}}  e^{-[(n+\frac{1}{2})b+(m+\frac{1}{2})/b]^2s}\label{imsp}
\ea
Closed string radiation for time-like tachyon condensation process was computed in \cite{Lambert:2003zr} \cite{Karczmarek:2003xm}. From the appendix of \cite{Lambert:2003zr} we can interpret the imaginary part of annulus diagram as the average total number of closed string emission which can be expressed as
\ba
\Im\mathbf{Z}=\bar{N}
\ea
where $\bar{N}$ is the average total number of closed string emission.

We will analyze the potential divergence of (\ref{imsp}) for open string IR divergence (large s) and UV divergence (small s). In the open string IR region, or equivalently $s \rightarrow \infty $ limit, $\eta(\frac{is}{2\pi})$ has the following asymptotic behavior
\ba
\eta(\frac{is}{2\pi})\sim \exp(-\frac{s}{24})
\ea
from which we have
\be \label{UVclosed}
\Im\mathbf{Z}\sim V_p\int^\infty \frac{ds}{s}\frac{1}{(4\pi s)^{\frac{p}{2}}}\exp\left(-\frac{s}{4b^2}+\frac{D-2}{24}s \right)
\ee
The exponent is negative as long as $b < \frac{1}{\sqrt{4-V^2}}$, which can be obtained with the help of $D+6V^2=26$. In our setup, $V^2>0$ is satisfied since we have set $V^+=0$ for convenience and taken the limit $b \rightarrow 0$. Combining the above conditions we have $b < \frac{1}{\sqrt{4-V^2}}$ and there is no IR divergence.
As for open string UV region, or equivalently $s \rightarrow 0 $ limit, $\eta(\frac{is}{2\pi})$ has the following asymptotic behavior
\ba \label{asp}
\eta(\frac{is}{2\pi})^{2-D}=(\frac{s}{2\pi})^{\frac{D-2}{2}} \exp[\frac{(D-2)\pi^2}{6s}] \left(1+(D-2)\exp(-\frac{4\pi^2}{s})+...\right)
\ea
where the ellipsis represents the high order term in $s^{-1}$.
From the asymptotic behavior (\ref{asp}) we have
\ba
\Im\mathbf{Z}\sim V_p\sum_{n,m=0}^{\infty}\int_0 \frac{ds}{s}\frac{1}{(4\pi s)^{\frac{p}{2}}}(\frac{s}{2\pi})^{\frac{D-2}{2}}\left[e^{\frac{D-2}{6s}\pi^2}+(D-2)e^{\frac{D-26}{6s}\pi^2}+... \right]e^{-[(n+\frac{1}{2})b+(m+\frac{1}{2})/b]^2s}\nonumber \\
\ea
The first term is from closed string tachyon which is uninteresting. Hence for $D<26$, there is no UV divergence in the open string channel as well.

\section{Open string pair production}
In this section, open string pair production in a linear dilaton background is calculated and its asymptotic behavior is analyzed.

In the process of unstable D-brane decay, in general there will be open string pair production on the brane because the hamiltonian of the system is time dependent. Some aspects of open string pair production on unstable brane have been discussed in \cite{Strominger:2002pc} \cite{Maloney:2003ck} \cite{Karczmarek:2003xm}. Open string pair production behavior implies that the in and out vacua are not the same. The vacuum decay amplitude is
\be \label{VA}
W=-\Re\ln\langle{\rm out} |{\rm in}\rangle
\ee
where the in and out vacuum are defined as
\be
a^{\rm in}|{\rm in} \rangle=0=a^{\rm out}|{\rm out} \rangle
\ee
Following the discussion of \cite{Karczmarek:2003xm}, we will work in an approximation in which the open strings propagate freely on the brane so that amplitude (\ref{VA}) is a one loop computation.
The in and out vacua are related by a Bogoliubov transformation
\ba
|{\rm out} \rangle=\prod_{k_{\parallel}}(1-|\gamma_{k_\parallel}|^2)^{1/4} e^{{-\frac{1}{2}\gamma^\ast_{k_\parallel}}(a_{k_\parallel}^{\rm in\dagger})^2}|{\rm in} \rangle
\ea
Using this relation, the vacuum decay amplitude can be written as
\be \label{vda}
\frac{W}{V_p}=-\frac{1}{4}\sum_{N}\int\frac{d^pk_\parallel}{(2\pi)^p}\ln(1-|\gamma_{k_\parallel}|^2)
\ee
where $\gamma_{k_\parallel}$ is the Bogoliubov coefficient and $k_\parallel$ is open string longitudinal spatial momentum which satisfies
\be
2\omega\omega'=k^2_\parallel +N-\frac{D-2}{24}
\ee
It is conjectured in \cite{Gutperle:2003xf} that the Bogoliubov coefficient $\gamma$ is related to the boundary two point function of dressed open string operators. Since we want to analyze the large $\omega$ behavior which is dominant, we can neglect the $\omega'$ dependence. At large $\omega$,
\be
|\gamma_\omega|^2 \sim \left|d(\omega+i\frac{V_+}{2})\right|^2
\ee
In order to calculate the boundary two point function of operator $e^{-i\omega X^+ +V_+X^+}$, we fix the zero modes of $X^i$ and $X^-$, and set $V_+=0$ for convenience.
The two-point function can be computed
\be
\langle e^{-i\omega X^+(z,\bar{z})} e^{-i\omega X^+(z',\bar{z}')} \rangle \sim \frac{\lambda^{\frac{2i\omega}{b}}}{b} \exp(V_-x^-)\Gamma(\frac{-2i\omega}{b})
\ee
The norm of two point function
\be \label{norm2p}
|d(\omega)|^2 \sim \frac{\pi\exp(2V_-x^-)}{2b\omega\sinh(\frac{2\pi\omega}{b})}
\ee
We would like to find the asymptotic behavior of (\ref{vda}) and (\ref{norm2p}) for large $\omega$. It follows that
\be
|d(\omega)|^2\sim \omega^{-1}e^{-\frac{2\pi\omega}{b}+O(\omega^0)}
\ee
The vacuum decay amplitude becomes
\be\label{opd}
\frac{W}{V_p}\sim \sum_{N}\int d^pk_\parallel e^{-\frac{2\pi\omega}{b}}\sim\int d\omega e^{2\pi\omega(\sqrt{\frac{D-2}{6}}-\frac{1}{b})}
\ee
The above exponent which describes the production of highly massive open string has the divergent behavior similar to (\ref{UVclosed}) which can be interpreted as the production of highly massive closed string. For $b <\sqrt{\frac{6}{D-2}}$, there is no divergence in (\ref{opd}), and for $b >\sqrt{\frac{6}{D-2}}$, there will be divergence in (\ref{opd}). Since $D+6V^2=26$, the above open string asymptotic behavior is consistent with the asymptotic behavior (\ref{UVclosed}) in the closed string channel. While deriving the asymptotic behavior (\ref{UVclosed}), we do need to take the limit $b\rightarrow 0$ as well as $b\omega\rightarrow 0$. Hence considering the above consistence, it is natural to expect the result (\ref{UVclosed}) to extend beyond the limit $b\rightarrow 0$ and $b\omega\rightarrow 0$.

\section{Discussion and Conclusion }

In this note we discuss the closed string emission from the light-like tachyon condensation process which represents the inhomogeneous decay of unstable D-brane. We show that the D-brane decay rate is UV finite, which gives some new insights into the light-like tachyon condensation. In \cite{Hellerman:2008wp}, Hellerman and Schnabl found a surprising difference in dynamics between the time-like tachyon condensation and the light-like tachyon condensation in the framework of the classical open string field theory. In the light-like case, the evolution of tachyon is given by gradient flow rather than Hamiltonian dynamics as in the time-like case. The gradient flow forces the tachyon to asymptote to the true vacuum at late times. As a result, the tachyon matter problem does not exist in the light-like tachyon condensation case. This raise the question of the quantum correction to the above picture, which is one of the motivations for this note. The intuitive picture is that the D-brane decays into closed strings which then run away rather than is localized in the D-brane's location as in the time-like case. The result that the open string one-loop correction is finite shows that the the classical picture of Hellerman and Schnabl remains valid under quantum correction and the perturbation theory is reliable. That is to say the unstable D-brane decays into closed strings before we are confronted with the strongly coupled theory. \\

We calculate the worldsheet partition function in the closed string channel using the boundary state/$\sigma$-model method. In some region of parameter space, we find that the D-brane decay rate is finite in the open string UV region and depends on tachyon profile in the open string IR region. The physical meaning of $b\rightarrow0$ is that the tachyon profile $\cT=\exp bX^+$ varies slowly along the $X^+$ direction compared to itself. Recall that if we recover the string length scale which has been set to one, we have $\cT=\exp bX^+/\sqrt{\alpha'}$. With these in hand, we can rewrite $b\omega\rightarrow0 $ as $\partial_+\cT/\cT\ll 1/\omega\alpha'$, which implies that in the emitted closed string scale the tachyon variation rate is small compared to itself. Roughly speaking, the working limit in this note requires that the tachyon varies slowly along $X^+$ direction compared to itself. In our setup the condition $b\ll1$ is essentially different from $b=0$ since the marginal condition $bV^+=1$ forbids us to take the limit $b=0$. So in the light-like tachyon case, we can not obtain the constant dilaton background from the linear dilaton background by adjusting the tachyon profile. This is another difference from the time-like case. We arrive at a conclusion about closed string emission when taking the limit as $b\rightarrow0$ and $b\omega\rightarrow0 $. It is interesting to compute the imaginary part of partition function beyond this limit. \\

We also compute the open string pair production which has similar asymptotic behavior to the closed string radiation. This provides further evidence to the open string completeness conjecture proposed in \cite{Sen:2003bc} \cite{Sen:2003xs}, which suggests that the closed string channel calculation provides a dual description of the open string channel calculation. In the light-like tachyon condensation process the underlying conformal field theory on the string worldsheet is some cousin of boundary Toda theory which has richer content than Liouville theory. However boundary Toda theory \cite{Fateev:2007ab} \cite{Fateev:2008bm} has not been well explored and needs to be paid more attention in the future. We expect that the one-point function calculation of this boundary Toda cousin in the appendix will give some insights to it.

\section*{Acknowledgments}
I have benefited from discussions with Bin chen, Wei He, Liang Kong, Chang-Yong Liu, Fei Liu, Jian-Feng Wu, Wei-shui Xu, Fei Ye, Peng Ye. I am grateful to Bin Chen for comments on the manuscript.

\appendix

\section{Exact bulk one-point function}

When we calculate open string partition function in section 3, we mainly focus on a two-dimensional conformal field theory whose action is
\ba \label{aa1}
S^\pm=-\frac{1}{2\pi}\int_\Sigma d^2\sigma\sqrt{g}g^{\alpha\beta}\partial_\alpha X^+\partial_\beta X^-+\frac{\lambda}{2\pi}\int_{\partial\Sigma}ds e^{b X^+}
\ea
The $X$ correlator on the unit disc is
\be
\left\langle X^\mu(z,\bar{z}) X^\nu(z',\bar{z}') \right\rangle=-\frac{\eta^{\mu\nu}}{2}\left[\ln|z-z'|^2 + \ln|1-z\bar{z}'|^2 \right])
\ee
where $\mu,\nu=+,-$ and the nonvanishing components $\eta^{+-}=\eta^{-+}=-1$.
The action (\ref{aa1}) can be rewritten as
\ba \label{aa2}
S^\pm=-\frac{1}{4\pi}\int_\Sigma d^2\sigma\sqrt{g}g^{\alpha\beta}\left(\partial_\alpha X^0\partial_\beta X^0-\partial_\alpha X^1\partial_\beta X^1\right)+\frac{\lambda}{2\pi}\int_{\partial\Sigma}ds e^{\frac{b}{\sqrt{2}} (X^0+X^1)}
\ea
This conformal field theory is some cousin of boundary Toda theory which has not been well explored.
In the subsection 3.1, we apply the technique developed by Hellerman and Schnabl in \cite{Hellerman:2008wp} to carry out an approximate calculation of bulk one-point function in which we fix the zero mode of field $X$ rather than integrate it out. In  this appendix, we intend to calculate the bulk one-point function exactly. Though the exact bulk one-point function is too complicated to be used in the brane decay calculation, it is heuristic and has its own right in the conformal field theory. Following the procedure developed in \cite{Goulian:1990qr} \cite{Gutperle:2003xf}, we will calculate the one-point function of closed string vertex operator $V=e^{\alpha_+X^+ +\alpha_-X^-}$ on the unit disc where $\alpha_+=V_+-i\omega$ and $\alpha_-=V_--i\omega'$. We decompose $X=x+\hat{X}$ where $x$ is the zero mode of $X$, and  $\hat{X}$ is the oscillation part of $X$. Under this decomposition, the one-point function on the upper half-plane becomes
\ba \label{eopf}
\langle V\rangle=
&\int dx^+dx^- D\hat{X}^+D\hat{X}^- e^{\frac{1}{2\pi}\int_{\Sigma}d^2z(\partial\hat{X}^+\bar{\partial}\hat{X}^-+ \partial\hat{X}^-\bar{\partial} \hat{X}^+)-\frac{\lambda e^{bx^+}}{2\pi}\int_{\partial\Sigma}dte^{b\hat{X}^+}}&  \nonumber\\ &\times e^{\alpha_+x^++\alpha_-x^-+\alpha_+\hat{X}^+ +\alpha_-\hat{X}^-}&
\ea
After the zero modes are integrated out, (\ref{eopf}) becomes
\ba
\langle V\rangle&=&
\frac{2\pi\delta(-i\alpha_- )(\frac{\lambda}{2\pi})^{-\frac{\alpha_+}{b}}\Gamma(\frac{\alpha_+}{b})}{b}\langle e^{\alpha_+\hat{X}^+ +\alpha_-\hat{X}^-}(z,\bar{z})\prod_{i=1}^n\int dt_ie^{b\hat{X^+}}(t_i)  \rangle_{\rm F} \label{eopf2}\\
&=&\frac{2\pi\delta(-i\alpha_- )(\frac{\lambda}{2\pi})^{-\frac{\alpha_+}{b}}\Gamma(\frac{\alpha_+}{b})}{b} |z-\bar{z}|^{\alpha_+\alpha_-} \prod_{i=1}^n\int dt_i |z-t_i|^{b\alpha_-}|\bar{z}-t_i|^{b\alpha_-}
\ea
where the subscript F denotes that the vacuum expectation value is calculated for the free part of the action.
About the formula (\ref{eopf2}), we have assumed $-\frac{\alpha_+}{b}=n$ to be some integer, and the general correlator is then given by analytic continuation in $n$. In order to calculate the integral, we transform the coordinate from the up half plane to the unit disc by conformal transformation
\ba
\omega=\frac{t-z}{t-\bar{z}};~~~~ t=\frac{\omega\bar{z}-z}{\omega-1}
\ea
In the unit disc coordinate, the one-point function of closed string vertex operator becomes
\ba
\langle V\rangle&=&\frac{2\pi\delta(-i\alpha_- )(\frac{\lambda}{2\pi})^{-\frac{\alpha_+}{b}}\Gamma(\frac{\alpha_+}{b})}{b} |z-\bar{z}|^{\alpha_+\alpha_-} \prod_{i=1}^n\int d\omega_i \frac{(z-\bar{z})|z-\bar{z}|^{2b\alpha_-}}{(\omega_i-1)^2|\omega_i-1|^{2b\alpha_-}}\nonumber \\
\ea
The integration part can be calculated directly
\ba
\langle V\rangle&=&\frac{2\pi\delta(-i\alpha_- )\Gamma(\frac{\alpha_+}{b})}{b} \left(\frac{\lambda \Gamma(1+2i\omega'b)}{(1+2i\omega'b)\Gamma(1+i\omega'b)\Gamma(i\omega'b)} \right)^{-\frac{\alpha_+}{b}} |z-\bar{z}|^{-\frac{\alpha_+}{b}-\alpha_+\alpha_-}\nonumber\\
\ea
As a result the exact bulk one-point function is
\ba
U(\alpha)=\frac{2\pi\delta(-i\alpha_- )\Gamma(\frac{\alpha_+}{b})}{b} \left(\frac{\lambda \Gamma(1+2i\omega'b)}{(1+2i\omega'b)\Gamma(1+i\omega'b)\Gamma(i\omega'b)} \right)^{-\frac{\alpha_+}{b}}
\ea

\end{document}